\documentclass[%
preprint,%
 amssymb, amsmath,%
pre,
floatfix
]{revtex4-1}

\usepackage[pdftex]{graphicx,color}
\usepackage{bm}%
\usepackage{ulem}%
\expandafter\ifx\csname package@font\endcsname\relax\else
 \expandafter\expandafter
 \expandafter\usepackage
 \expandafter\expandafter
 \expandafter{\csname package@font\endcsname}%
\fi
\begin{document}

\preprint{aaa/bbb}

\title{Active lift inversion process of heaving wing in uniform flow by temporal change of wing kinematics}%

\author{Makoto Iima}%
	\email{iima@hiroshima-u.ac.jp}
	\affiliation{Graduate School of Science, Hiroshima University,\\ 
1-7-1, Kagamiyama Higashi-Hiroshima, Hiroshima 749-8251, Japan}

\author{Naoto Yokoyama}%
	\affiliation{
Department of Mechanical Science and Bioengineering, Osaka University,\\
1-3 Machikaneyama, Toyonaka, Osaka 560-8531, Japan}
	
\author{Kei Senda}%
	\affiliation{Graduate School of Engineering, Kyoto University, 
C3 Building, Kyoto Daigaku-Katsura, Nishikyo-ku, Kyoto 615-8540, Japan}

\date{\today}%

\begin{abstract}
The transition of the vortex pattern and the lift generated by a heaving wing in a uniform flow was investigated numerically. 
As a fundamental problem constituting the insects' flight maneuverability, we studied the relationship between a temporal change in the heaving wing motion and the change in the global vortex pattern. At a Strouhal number that generates an asymmetric vortex pattern, we found that temporal angular frequency reduction causes inversion of both the global vortex pattern and the lift sign. The inversion is initiated by the transfer of the leading-edge vortex, which interferes with the vortex pattern generated at the trailing edge. Successful inversion is conditioned on the starting phase and the time interval of the frequency reduction. The details of the process during the transition are discussed. 
\end{abstract}

\pacs{aaa}

\maketitle
\section{Introduction}
\vspace{-6pt}
Animal locomotion in fluids, such as the flight of insects and the swimming of fish, is achieved by the unsteady flow driven by wings or fins. During flight, insects exploit vortex structures generated by the motion of their wings, which makes the flight mechanisms different from those of conventional fixed-wing aerodynamics. Such mechanisms include delayed stall, rotational circulation, the clap-and-fling mechanism, and wing-wake interaction
\cite{chin16_flapp_wing_aerod};
 several reviews are available
\cite{%
ellington84_aerod_hover_insec_flighIV,%
sane03_aerod_insec_fligh,%
wang05_dissec_insec_fligh,%
chin16_flapp_wing_aerod%
}.
When the flight (or swimming) is steady (time-periodic), the generation, transfer, merging, and dissipation of the vortex structure during one flapping cycle are periodic.

Real insects need to maneuver their flight, i.e., control their flight speed and/or direction, e.g., take-off, landing, and changing speeds of forward flight\cite{%
Betts-Wootton1988Wing,%
Dudley2000TheBiomechanics%
}.
Consequently, their flight is unsteady (non-periodic), and the flight mechanisms or generated vortex structures can change. 
For instance, butterflies fly with a sequence of several flight modes and rapid maneuvers
\cite{Betts-Wootton1988Wing},
 and they uses a variety of flight mechanisms in successive strokes
\cite{srygley02_uncon_lift_gener_mechan_free_flyin_butter}.
Such changes of flight mechanisms require the control of particular vortex structures via wing kinematics.
Although flight maneuverability is an important aspect of flapping flight, the study of maneuverability is limited to observations
\cite{srygley02_uncon_lift_gener_mechan_free_flyin_butter,%
lin12_signif_momen_of_inert_variat,%
hedrick07_low_speed_maneuv_fligh_rose%
} or numerical simulations\cite{%
ramamurti07_comput_inves_three_dimen_unstead,%
fei16_impor_body_rotat_durin_fligh_butter%
}
 of real animals.

For flight maneuverability and stability, unsteady wing and/or body motion are required to maintain the flight, which is regarded as unstable in many studies\cite{%
sun05_high_lift_gener_power_requir_insec_fligh,%
taylor05_nonlin_time_period_model_longit,%
faruque10_dipter_insec_fligh_dynam,%
senda12_model_emerg_flapp_fligh_butter,%
yokoyama13_aerod_forces_vortic_struc_flapp%
}. 
The relationship between such wing and/or body motion and the related fluid dynamics has not been clarified, except for a ``damping factor'' highlighted in studies on maneuverability
\cite{hedrick11_dampin_flapp_fligh_its_implic}
 and stability
\cite{gao11_pertur_analy_fligh_dynam_passiv}.

Despite the fundamental importance of actively exploiting vortex structures, to the best of the authors' knowledge, few studies have investigated the fluid dynamics that connects the vortex structures, including lift generation and unsteady (non-periodic) wing kinematics. This is partly because of the nature of the unsteady flight, i.e., a strongly-coupled system consisting of (1) the motion of the center of mass and the orientation of the body, (2) body and wing motion (deformation of the animal's body), and (3) the motion of the surrounding fluid, even if we omit the sensing and control parts. Clearly, the entire system is too complex to be resolved all at once.
We should break up the entire problem into fundamental pieces easier to handle with, which will contribute to resolve the maneuver problem.

Here, we remark that the analysis of the coupled system comprising (2) and (3) is interesting and difficult by itself as a fluid mechanics problem because of the non-periodicity and the strongly nonlinear nature of fluid dynamics. Flow dynamics due to non-periodic wing motion has not been studied in detail, except for an impulsively started object as the simplest example
\cite{%
taneda71_unstead_flow_past_flat_plate,%
huang01_surfac_flow_vortex_shedd_impul_start_wing,%
ringuette07_role_tip_vortex_force_gener,%
taira09_three_dimen_flows_aroun_low%
}.
In this study, we focus on the relationship between vortex patterns (and the associated hydrodynamic force) and wing kinematics, especially for the effect of non-periodic wing motion. If the hydrodynamic force vector and the vortex structure can be controlled via wing motion, such wing motion will be of potential use for efficient flight or propulsion control, e.g., control without additional apparatus such as a flap or rudder. 
Because the description of non-periodic wing motion requires many parameters, we consider a model with simple wing kinematics, i.e., a heaving wing in a uniform flow with the given wing kinematics, to highlight the intrinsic nature of the wing-vortex interaction, though the model of insects' flapping motion should include other kinematics such as flapping motion.

Various types of vortex patterns are generated by an oscillating wing. Several studies have investigated vortex patterns generated by a heaving wing in a uniform flow\cite{%
jones98_exper_comput_inves_knoll_betz_effec,%
Parker2001,%
lewin03_model_thrus_gener_two_dimen,%
ellenrieder07_piv_measur_asymm_wake_two,%
heathcote07_jet_switc_phenom_period_plung_airfoil,%
lua07_wake_struc_format_heavin_two,%
jones09_desig_devel_consid_biolog_inspir,%
michelin09_reson_propul_perfor_heavin_flexib_wing%
}. 
In particular, wake deflection is an asymmetric vortex pattern that is ubiquitous when both the Strouhal number and the heaving amplitude are large. In this case, the sign of the average lift depends on the direction of the wake. Wake deflection has been experimentally observed in the case of both high-aspect-ratio wings\cite{%
jones98_exper_comput_inves_knoll_betz_effec,%
ellenrieder07_piv_measur_asymm_wake_two%
}
 and low-aspect-ratio wings\cite{%
Parker2001%
}.
The wake direction (deflection angle) is constant or time dependent\cite{%
jones09_desig_devel_consid_biolog_inspir,%
jones98_exper_comput_inves_knoll_betz_effec,%
heathcote07_jet_switc_phenom_period_plung_airfoil%
}.
Similar asymmetric vortex patterns have been reported for a simple flapping wing\cite{%
spagnolie10_surpr_behav_flapp_locom_with_passiv_pitch,%
schnipper09_vortex_wakes_flapp_foil,%
godoy-diana09_model_symmet_break_rever_benar,%
godoy-diana08_trans_wake_flapp_foil%
}, a wing with both heaving and flapping motion\cite{%
michelin09_reson_propul_perfor_heavin_flexib_wing,%
muijres07_wake_visual_heavin_pitch_foil_soap_film,%
ellenrieder03_flow_struc_behin_heavin_pitch,%
bose18_inves_chaot_wake_dynam_past%
}, and even for a wing model that can move according to the generated thrust\cite{%
vandenberghe06_unidir_fligh_free_flapp_wing,%
alben05_coher_locom_as_attrac_state,%
zhang10_locom_passiv_flapp_flat_plate,%
shelley11_flapp_bendin_bodies_inter_with_fluid_flows%
}.

In this paper, we use a simple model to show that temporal frequency reduction can cause inversion of the deflected wake pattern. The parameters are chosen such that the non-dimensional parameters are set in the range of insects\cite{%
sato07_strok_frequen_but_not_swimm,%
taylor03_flyin_swimm_animal_cruis_at%
}. 
By limiting the wing kinematics, we clarify the parameter region for the non-trivial vortex transition. A previous study has reported vortex pattern transitions when flapping is abruptly stopped in the case of a two-dimensional free-flight model\cite{Iima2013AST}; however, owing to the coupling between the vortex dynamics and the center-of-mass motion, the separation dynamics and the vortex dynamics in the parameter space of wing kinematics were not examined in detail. The model analyzed here is simplified considerably to focus on changes in the vortex pattern on the basis of smooth wing kinematics.

The remainder of this paper is organized as follows. Sec. \ref{sec:Model} describes the details of the model and the numerical method. Sec. \ref{sec:Results} presents the results. Vortex structures and the lift and drag in simple heaving motion are discussed in Sec. \ref{sec: Simple heaving},
 whereas the transition of the vortex structures owing to non-periodic wing motion is discussed in Sec. \ref{sec:Temporal reduction of heaving frequency}. 
Further, the Reynolds number dependence on the discovered vortex transition process is discussed in Sec. \ref{sec: Discussion: The Reynolds number dependency}. 
Finally, Sec. \ref{sec:Concluding Remarks} concludes the paper.

\section{Model}
\label{sec:Model}
\subsection{Wing kinematics}
\begin{figure}[h]
\begin{center}
\includegraphics[width=0.4\textwidth]{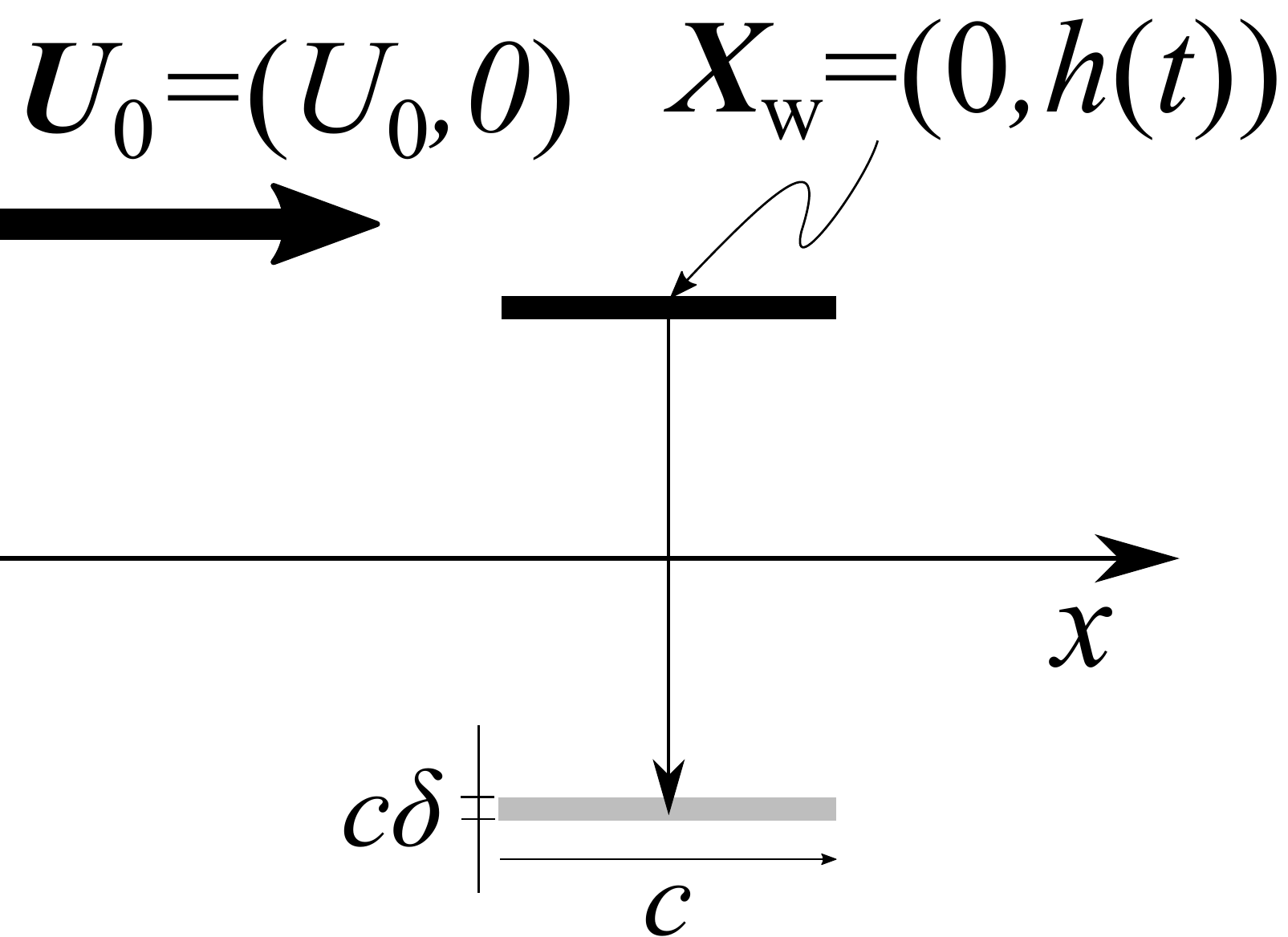}
\end{center}
\caption{
Configuration of the model. A wing in a two-dimensional uniform flow oscillates perpendicular to the uniform flow. The position of the center of the wing is described by $\bm{X}_w=(0, h(t))$.
}
\label{fig:model}
\end{figure}
A heaving wing in a two-dimensional uniform flow (Fig. \ref{fig:model}) is considered. We assume that the wing is a plate with wing chord length $c$ and thickness $c\delta$, and that both ends are semi-circles of radius $c\delta /2$. The wing, which oscillates vertically to the uniform flow $\bm{U}_0=(U_0,0)$, is always parallel to $\bm{U}_0$, and the center of the wing, $\bm{X}_w$, moves perpendicular to $\bm{U}_0$: $\bm{X}_w=(0,h(t))$. The function $h(t)$ is defined as 
\begin{equation}
h(t) = A \sin \Phi(t),
\label{eq:function of heaving 1}
\end{equation}
where the constant $A$ is the oscillation amplitude and the function $\Phi(t)$ is the phase of the oscillation, which determines the details of the wing motion.
When $\Phi(t)=\omega t$ ($\omega$ is the constant angular velocity), the wing motion is simple heaving. To describe the general wing kinematics, we need an infinite number of parameters. However, in this paper, the wing motion is restricted such that $\Phi(t)$ is described as follows: 
\begin{eqnarray}
\Phi(t)=\Phi(t; \omega, \Delta \omega, t_1, t_2)
 &=&
 \phi + \omega t - (t_2-t_1) F(t; t_1, t_2) \Delta \omega, \label{eq:function of phase}
\end{eqnarray}
where the constants $\phi$ and $\Delta \omega$ are the initial phase of oscillation and the decrement in angular frequency, respectively. The function $F(t;t_1,t_2)$ is defined as 
\begin{eqnarray}
  F(t; t_1, t_2) &=& \sigma \left( t-\dfrac{t_1+t_2}{2},\; \dfrac{4}{t_2-t_1} \right),\quad 
 \sigma(x,a) = \dfrac12 \left( \tanh\dfrac{ax}{2} + 1 \right). \label{eq:sigmoid}
\end{eqnarray}
where $\sigma(x,a)$ is the sigmoid function that connects 0 and 1 smoothly around $x=0$, i.e.,  $\displaystyle \lim_{x \to -\infty} \sigma(x,a)=0$ and $\lim_{x \to \infty} \sigma(x,a)=1$, with characteristic width $1/a$. Thus, the function $\Phi(t; \omega, \Delta \omega, t_1, t_2)$ shifts the phase of oscillation by $- \Delta \omega (t_2-t_1)$; in other words, the local angular velocity, defined by $\displaystyle \partial \Phi/\partial t$, undergoes a temporal decrease of $-\Delta \omega$ at $t=(t_1+t_2)/2$. The change mainly occurs in the time interval $[t_1,t_2]$ (cf. Fig. \ref{fig:wing motion}).
\subsection{Numerical method of fluid motion}
\begin{figure}[h]
\begin{center}
\includegraphics[width=0.9\textwidth]{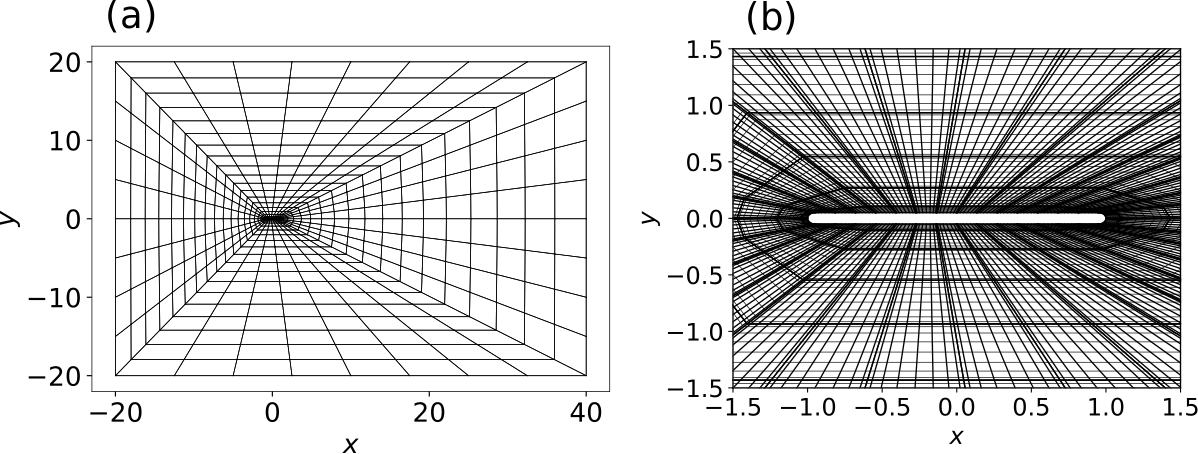}
\end{center}
\caption{
(a) Sub-regions for the simulation domain.
(b) Sub-regions (divided by thick lines) and the collocation points (represented by the crossing points of thin lines) near the wing.
}
\label{fig:Meshes}
\end{figure}

To numerically solve the fluid motion according to the wing motion, we use a coordinate system in which the wing is fixed. Let us define the coordinate system in the laboratory frame as $O-xy$ and the wing-fixed coordinate system as $O-XY$.

The velocity and acceleration of the center of the wing in the $O-xy$ system are
$\bm{V}_w = (0, h'(t))$
 and
$\bm{A}_w = (0, h''(t))$, respectively. When the fluid velocity in the $O-xy$ system is $\bm{u}$, the fluid velocity in the $O-XY$ system, $\bm{U}$, is given by
 $\bm{U}=\bm{u}-\bm{V}_w$. 
Similarly, when the force in the $O-xy$ system is $\bm{f}$, the force acting on the wing in the $O-XY$ system, $\bm{F}$, is given by $\bm{F} = \bm{f}-\rho_w B \bm{A}_w$. 
Note that the force calculated in the $O-XY$ system includes an artificial force proportional to the acceleration and volume of the wing; $\rho_w B \bm{A}_w$, where $\rho_w$ and $B$ are the density and volume (area in the two-dimensional case) of the wing, respectively. These formulae give the transformation between variables in both the coordinate systems. In the following, we represent the values of all the variables in the $O-xy$ coordinates, although the calculation was performed in the $O-XY$ coordinates (for comparison with the calculation using the immersed boundary method in the laboratory frame, refer to the Appendix).

The fluid motion is governed by the incompressible Navier–Stokes (NS) equations: 
\begin{equation}
	\dfrac{\partial \bm{u}}{\partial t}
+ \bm{u}\cdot \nabla \bm{u} 
 = - \dfrac{1}{\rho}\nabla p
	 + \nu \nabla^2 \mbox{\boldmath$u$}, \;\; \nabla \cdot \bm{u}=0,
	 \label{eq:NS}
\end{equation}
where $\mbox{\boldmath$u$}=(u,v)$ is the velocity field, $p$ is the pressure, and $\nu$ is the kinematic viscosity.

In this model, we have three non-dimensional parameters:
Reynolds number $\displaystyle Re=\dfrac{U_0 c}{\nu}$,
Strouhal number $\displaystyle St=\dfrac{fA}{U_0}$ ($\displaystyle f=\dfrac{\omega}{2\pi}$),
and the non-dimensional amplitude $\displaystyle r=\dfrac{A}{c}$.

The fluid motion is solved using the spectral element method (SEM), which is a high-order finite element technique that combines the geometric flexibility of finite elements with the high accuracy of spectral methods. We used {\it Semtex} \cite{blackburn04_formul_galer_spect_elemen_fourier},
 an open-source SEM package that has been used in many hydrodynamic problems.

In this study, the computational domain is $[X_1, X_2] \times [Y_1, Y_2]$. The boundary condition at the outer sides of the domain is inflow with $\bm{u}=\bm{U}_0$ except for the right side ($x=X_2, Y_1 \le y \le Y_2$), where the robust outflow condition proposed by Dong et al.\cite{dong14_robus_accur_outfl_bound_condit}
 with a smoothness parameter (``$\delta$'' in their paper) of 0.1 is applied. The domain is decomposed into $N_1 \times N_2$ quadrilateral sub-regions (``elements'' in the Semtex manual) with $O-$type geometry; $N_1$ and $N_2$ denote the number of divisions in the azimuthal and radial directions, respectively. Each sub-region contains $M^2$ Lagrange knot points (Fig. \ref{fig:Meshes}).

The parameters of the system were as follows: 
$U_0=1, \rho=1, c=2, \delta=0.05, A=0.6$, which gives and $r=0.3$.
For numerical simulation, we used $N_1=32, N_2=20, M=9, X_1=-20,X_2=40, Y_1=-20,Y_2=20$. The time integration was of the second order with time step $\Delta t=1.0 \times 10^{-4}T$, where $T=1/f$ is the heaving period. The grid width in the sub-regions attached to the wing ranged from $2.13\times 10^{-3}$ to $2.83 \times 10^{-2}$. 
The initial state was $\bm{u}(t=0)=\bm{U}_0$.
In the following, $Re=200$ and we controlled $St, t_1, t_2$ and $\Delta \omega$
 except for Sec. \ref{sec: Discussion: The Reynolds number dependency} where $Re$ was changed.

We confirmed that the main result with the physical parameters $(St, \Delta \omega, t_1, t_2)=(0.275, \omega/2, 7 \dfrac13, 8\dfrac13)$ (see Sec. \ref{sec:Temporal reduction of heaving frequency})
 as well as the results with the simple heaving ($\Delta \omega=0$, $0.1 \le St \le 0.3$)
 did not change with finer simulation parameters $(N_1,N_2,M, \Delta t)=(40, 25, 11, T/15000)$.
\section{Results}
\label{sec:Results}
\subsection{Simple heaving}
\label{sec: Simple heaving}
\subsubsection{Transition of vortex pattern}
\label{sec: Transition of vortex pattern}
\begin{figure}[h]
\includegraphics[width=0.9\textwidth]{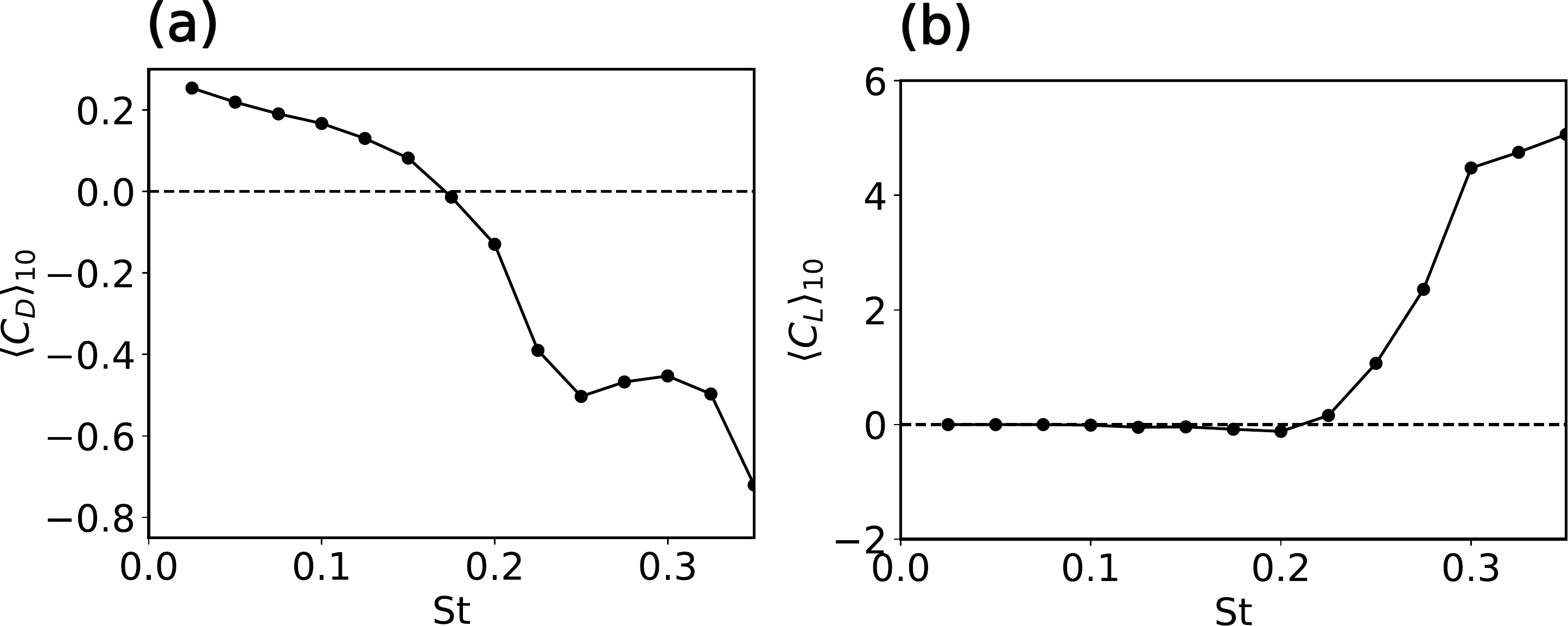}
\caption{
$\langle C_D \rangle_{10}$ and $\langle C_L \rangle_{10}$ for 
$\phi=\pi$.
(a): $\langle C_D \rangle_{10}$ vs. $St$.
(b): $\langle C_L \rangle_{10}$ vs. $St$.
}
\label{fig:CL and CD for simple heaving motion}
\end{figure}

\begin{figure}
\includegraphics[width=0.9\textwidth]{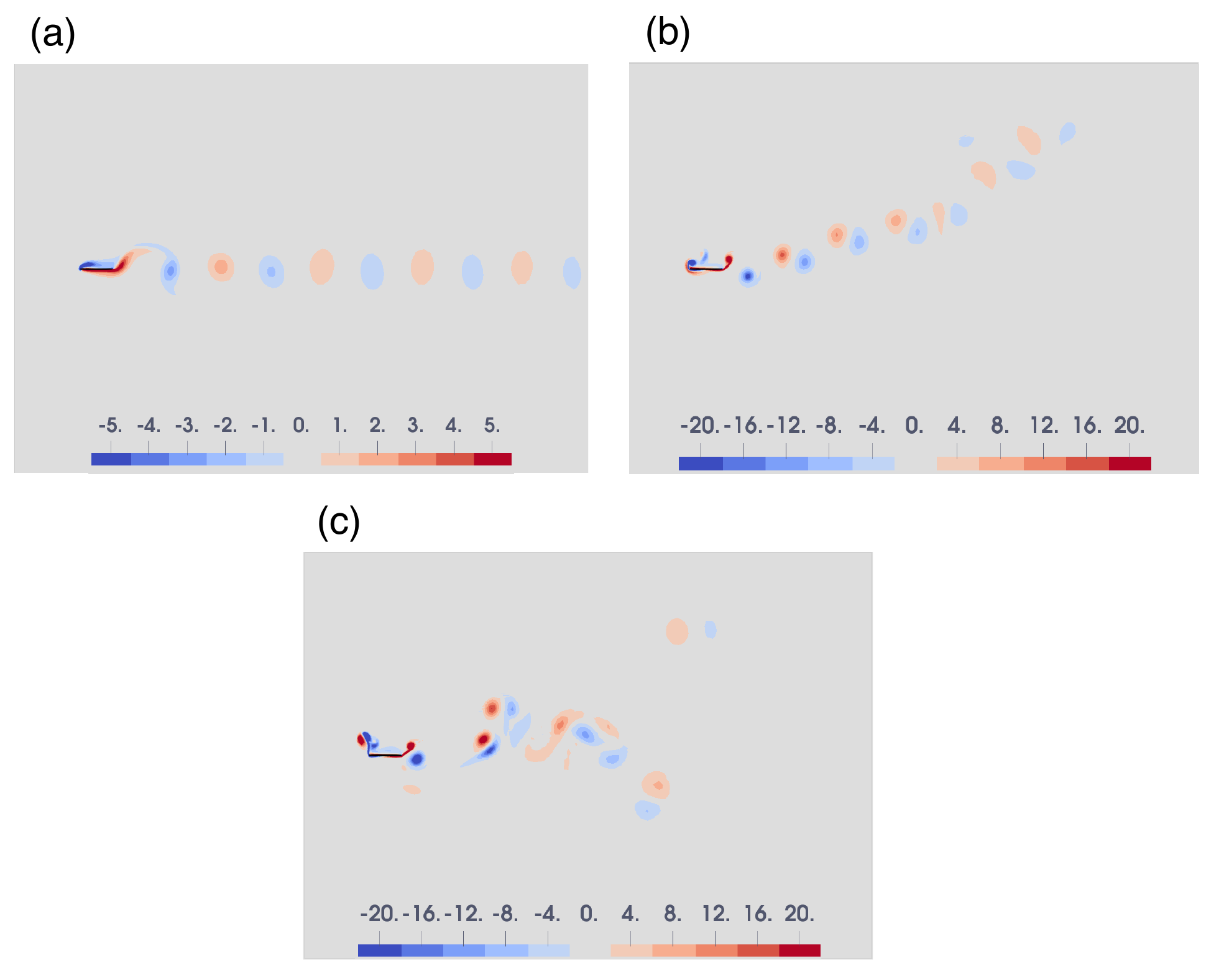}
\caption{
Vortex pattern at $t=10T, \phi=\pi$.
The displayed region is $[-5,30]\times[-12.5,12.5]$.
The colors indicate the vorticity, $\nabla \times \bm{u}$.
(a) $St=0.10$.
(b) $St=0.275$.
(c) $St=0.35$.
}
\label{fig:Vortex patterns for simple heaving}
\end{figure}
First, we show the results with $\Delta \omega=0$ (simple heaving) to explain the vortex patterns observed in this system. In this case, the heaving motion is periodic and the vortex patterns depend on $St$
\cite{jones98_exper_comput_inves_knoll_betz_effec,jones09_desig_devel_consid_biolog_inspir}. 
We considered the case of $\phi=\pi$
 for the integration time $10T$. 
Figure \ref{fig:CL and CD for simple heaving motion} shows
 $\langle C_L \rangle_{10}$ and $\langle C_D \rangle_{10}$ for $0.025 \le St \le 0.35$,
 where $\langle C_L \rangle_n$ and $\langle C_D \rangle_n$ denote the period-averaged lift and period-averaged drag, respectively, 
$\displaystyle
\langle C_L \rangle_n=\frac{1}{T}\int_{(n-1)T}^{nT} C_L dt
$, and a similar formula applies to $\langle C_D \rangle_n$.

Figure \ref{fig:CL and CD for simple heaving motion}(a) shows that the sign of $\langle C_D \rangle_{10}$ changes at $St\simeq 0.175$. 
When $St$ is below the critical value, the horizontal force acting on the wing is positive (drag), and a negative force (thrust) is generated above the critical value. 
Figure \ref{fig:CL and CD for simple heaving motion}(b) shows the transition of $\langle C_L \rangle_{10}$ from smaller values to order-of-unity values occurring at $St \simeq 0.20$,
 i.e., a transition from a symmetric vortex pattern to an asymmetric one. 
A symmetric vortex pattern with drag ($St=0.10$) is shown in Fig. \ref{fig:Vortex patterns for simple heaving}(a) and an asymmetric vortex pattern, i.e., wake deflection, with thrust ($St=0.275$) is shown in Fig. \ref{fig:Vortex patterns for simple heaving}(b). 
The major vortex structure is generated up to t=10T. The asymmetric vortex pattern loses its order when $St \ge 0.325$. 
In this case, both the leading-edge vortex (LEV) and the trailing-edge vortex (TEV) are released from the wing to generate an irregular pattern (Fig. \ref{fig:Vortex patterns for simple heaving}(c)).
A chaotic flow generation due to LEV-TEV interaction was analyzed in the case of the pitching and heaving wing with larger Reynolds number (1,000)
\cite{%
bose18_inves_chaot_wake_dynam_past%
}

\subsubsection{Vortex dynamics in wake deflection}
\label{sec: Vortex dynamics in wake deflection}
\begin{figure}[h]
\includegraphics[width=0.8\textwidth]{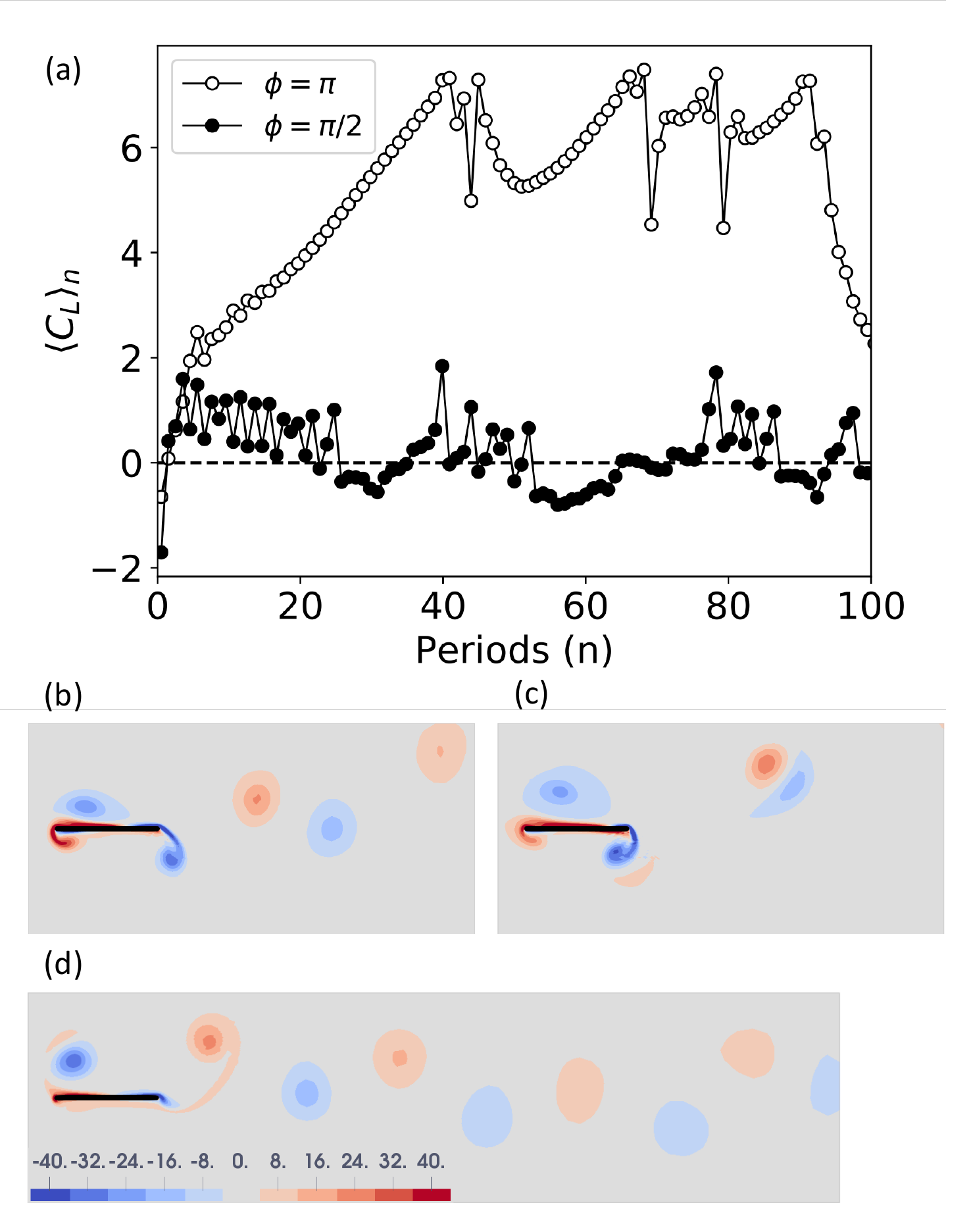}
\caption{
(a)
Plot of $(n-\dfrac12, \langle C_L \rangle_n)$ for different initial phases ($\phi=\pi/2, \pi$). 
(b) Snapshot of the wing and vortices for $t=10.5T$ and $\phi=\pi$.
 Color indicates vorticity; color boundaries are $\nabla \times \bm{u}=\pm (4+8k)\;(k=0,1,\cdots,5)$.
(c) Same as (b), but for $t=40.5T$.
(d) Same as (b), but for $t=70.5T$ and $\phi=\pi/2$.
}
\label{fig:CL for Long time}
\end{figure}

\begin{figure}
\includegraphics[width=0.8\textwidth]{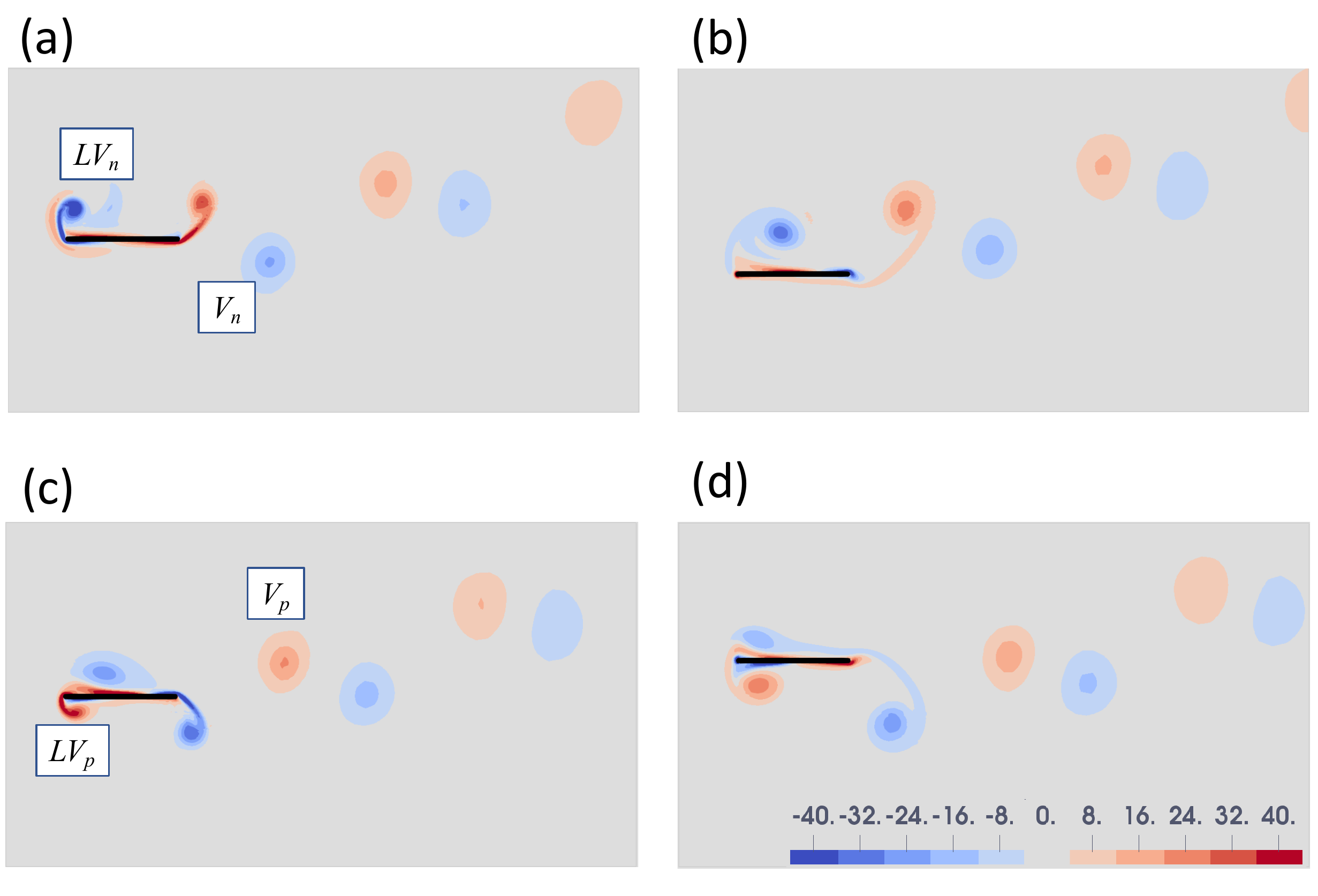}
\caption{
Time sequences of vortex patterns for $St=0.275$.
(a) $t=9.00T$.
(b) $t=9.25T$.
(c) $t=9.50T$.
(d) $t=9.75T$.
}
\label{fig:Leading-edge vortex for St0.275}
\end{figure}

\begin{figure}[h]
  \includegraphics[width=0.6\textwidth]{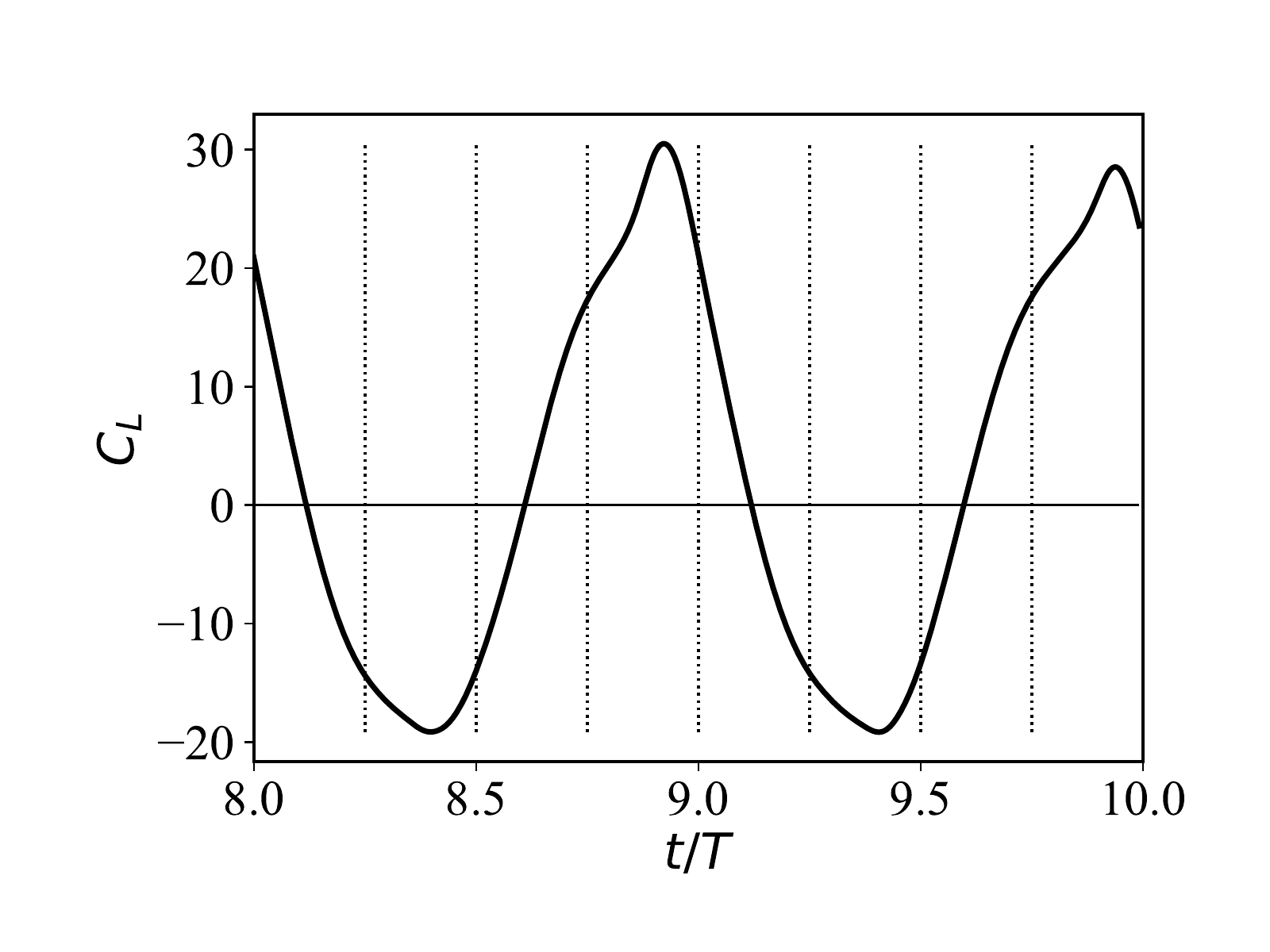}
  \caption{
$C_L$ vs. $t/T$ for $St=0.275$ ($8T \le t \le 10T$).
}
\label{fig:CL timeseries}
\end{figure}

The details of the asymmetric vortex pattern also depend on the initial phase $\phi$; two cases,
 $\phi=\pi/2$ and $\phi=\pi$,
 were compared.
In the range $St \le 0.20$ (symmetric vortex pattern), there is no significant difference between the two cases. However, a difference is observed when $St \ge 0.20$,
 which suggests that the asymmetric vortex pattern depends on $\phi$, though the symmetry is broken in both the cases.

In this paper, we define the asymmetric vortex pattern for large values of $\left|\langle C_L \rangle_n \right|$ (larger than 2.0), i.e., ``deflected wake'' \cite{jones98_exper_comput_inves_knoll_betz_effec}. In this case, the wake deflection is clear, and the deflection angle of the asymmetric pattern $\alpha$, i.e., the angle between the $x-$axis and the line passing near the trailing edge and the separation vortices, takes a positive value. The angle depends on the initial condition; the inverted pattern with the deflection angle $-\alpha$ can be obtained when the initial phase is shifted by $\pi$.

We remark that this difference of $\langle C_L \rangle_n$ for the initial phase $\phi$ is not transient. 
In Fig. \ref{fig:CL for Long time}(a), $\langle C_L \rangle_n$ for $\phi=\pi/2$ and $\phi=\pi$ for $St=0.275$ are shown for the number of periods, $n$. 
In the case of $\phi=\pi$, the deflection angle keeps positive (upward) from the beginning of wake formation, but the size of LEV becomes large and the wake structure changes (i.e. the distance between the vortices consisting of the dipoles becomes shorter) as $n$ becomes large
 (Fig. \ref{fig:CL for Long time}(b) and (c)).
Then, LEV interacts with the wake via TEV, which causes an instability of the whole vortex structure to fluctuate.
Such LEV-TEV interaction is also referred to in the context to a trigger to chaos in the flow around the heaving-pitching wing for larger Reynolds number
\cite{bose18_inves_chaot_wake_dynam_past}
.
On the other hand, in the case of $\phi = \pi / 2$, the vortex structure is horizontal rather than deflected (Fig. \ref{fig:CL for Long time}(d)).
The line passing between positive and negative vortices waves aperiodically as it goes to downstream, which causes small fluctuations of $\langle C_L \rangle_n$.
Clearly, $\langle C_L \rangle_n$ depends on $\phi$, and the difference is observed up to at least $100T$\footnote{
We remark that $C_L$ is defined by using $U$ as the typical velocity, while the wing speed includes the effect of heaving motion.
If we take the typical velocity $\textrm{max}\sqrt{U^2+ |\bm{V}_w|^2} =\sqrt{U^2+ (A \omega)^2}=U\sqrt{1+(2\pi St)^2}\simeq 2.00 U$, $C_L$ is reduced approximately $1/4$.
}.
Such simulations of the heaving wing suggest that the deflection angle varies with time over a long time scale (over 100 periods)\cite{heathcote07_jet_switc_phenom_period_plung_airfoil}, but the present integration time is not sufficiently long to diagnose the periodicity. 
It seems that such long-lasting initial phase dependence has not been reported.

The formation of the asymmetric vortex pattern is related to the values of $\langle C_L \rangle_n$. 
The interference of separation vortex generation with the vortex structure may be represented by the increment in $\langle C_L \rangle_n$, i.e., $\langle C_L \rangle_n-\langle C_L \rangle_{n-1}$ (or gradient of the graph). 
Figure \ref{fig:CL for Long time} shows that the increments in $\langle C_L \rangle_n$ for $n<7$ are larger than those for $7<n<40$, which suggests that the vortex structure generated before $n=7$ is qualitatively different. 
This observation is in agreement with the fact that the major vortex pattern is generated up to $10T$
 (Figs. \ref{fig:Leading-edge vortex for St0.275}(a) and (b)). 
The function $\langle C_L \rangle_n$ for $n>40$ shows fluctuations due to the stability of the deflected wake with many vortices. Hereafter,
we mainly focus on the vortex structure generated until around $t=15T$,
 which covers the critical number of periods determined by the lift increment.
Focusing on this time range, we can omit the effect of the boundary condition because the wake structure does not reach the outer boundary, and 
 the instability of the wake is not observed.
This time range also covers the minimum period for the lift inversion discussed in Sec. \ref{sec:Temporal reduction of heaving frequency}.

The maintenance of the deflected wake is due to the following factors: (1) generation of the asymmetric vortex pattern by the TEV, and (2) non-interference of the LEV with the TEV. These factors are clearly observed in Fig. \ref{fig:Leading-edge vortex for St0.275}, where the sequence of the vortex patterns near the wing is shown over one period (See Supplemental Material []Re200SimpleHeaving.mp4]).

For the first factor, $St$ is larger than those for symmetric vortex patterns, which means that the angular frequency and the heaving speed are also larger. Thus, the generated TEV has larger circulation. The interactions of the TEV with other shed vortices are stronger, and they break the symmetry of the vortex pattern with respect to the direction of uniform flow. This mechanism is reproduced by the simulation without the LEV, which was determined by the discrete vortex method considering the separation from the trailing edge alone \cite{jones98_exper_comput_inves_knoll_betz_effec}.

For the second factor, we follow the dynamics of the LEV. Two LEVs with positive and negative signs are generated during upstroke and downstroke, respectively. We focus on the LEV with the negative sign generated during the downstroke ($LV_n$ in Fig. \ref{fig:Leading-edge vortex for St0.275}(a)). The LEV is connected with the leading edge via a thin vortex layer (Fig. \ref{fig:Leading-edge vortex for St0.275}(b)) before it is stretched and dissipated owing to the upstroke (Figs. \ref{fig:Leading-edge vortex for St0.275}(c) and (d)). 
However, a part of the vortex remains and merges with the separation vortex generated during the next downstroke. On the other hand, the LEV with the positive sign generated during the upstroke ($LV_p$ in Fig. \ref{fig:Leading-edge vortex for St0.275}(c)) is stretched and most of it is dissipated. These processes show that the LEV does not interfere with the TEV dynamics significantly.

Neither dissipation nor trapping of LEVs occur when $St$ is much larger ($St \ge 0.35$); the LEVs also detach from the wing and disturb the patterns due to the TEVs, and the entire vortex pattern becomes irregular (Fig. \ref{fig:Leading-edge vortex for St0.275}(c)). In other words, the suppression of LEV-TEV interference gives an important condition for generation of the deflected wake. Controlling the LEV-TEV interference might enable us to realize a change in the vortex structure.

The instantaneous lift coefficient is shown in Fig. \ref{fig:CL timeseries}. Asymmetric lift generation is clearly observed. In the interval $9T < t < 10T$, positive lift is generated when $t < 9.12T$ and $9.60T < t$, whereas negative lift is generated when $9.12T < t < 9.60T$. 
As $\phi=\pi$, the downstroke is observed when $t < 9.25T$ and $9.75T< t$,
 which corresponds to the interval of positive lift in an approximate sense. The vortex dynamics in the positive lift generation is shown in Figs. \ref{fig:Leading-edge vortex for St0.275}(a) and (d) (the pattern in Fig. \ref{fig:Leading-edge vortex for St0.275}(d) is nearly the same as the pattern at $t=8.75T$). 
In this sequence, the free vortex with a negative sign near the TEV ($V_n$ in Fig. \ref{fig:Leading-edge vortex for St0.275}(a)), which was detached from the trailing edge, remains near the trailing edge; thus, the lift generation is enhanced. On the other hand, the negative lift generation is relatively weak because the free vortex with a positive sign near the TEV ($V_p$ in Fig. \ref{fig:Leading-edge vortex for St0.275}(c)) is not as close to the trailing edge as in the downstroke. The effect of the free vortex on the lift generation is reminiscent of the wake capture in insect flight\cite{dickinson99_wing_rotat_aerod_basis_insec_fligh}. 
The asymmetric lift generation results in a non-zero value of the total lift.

\subsection{Temporal reduction of heaving frequency}
\label{sec:Temporal reduction of heaving frequency}
\begin{figure}
\includegraphics[width=0.8\textwidth]{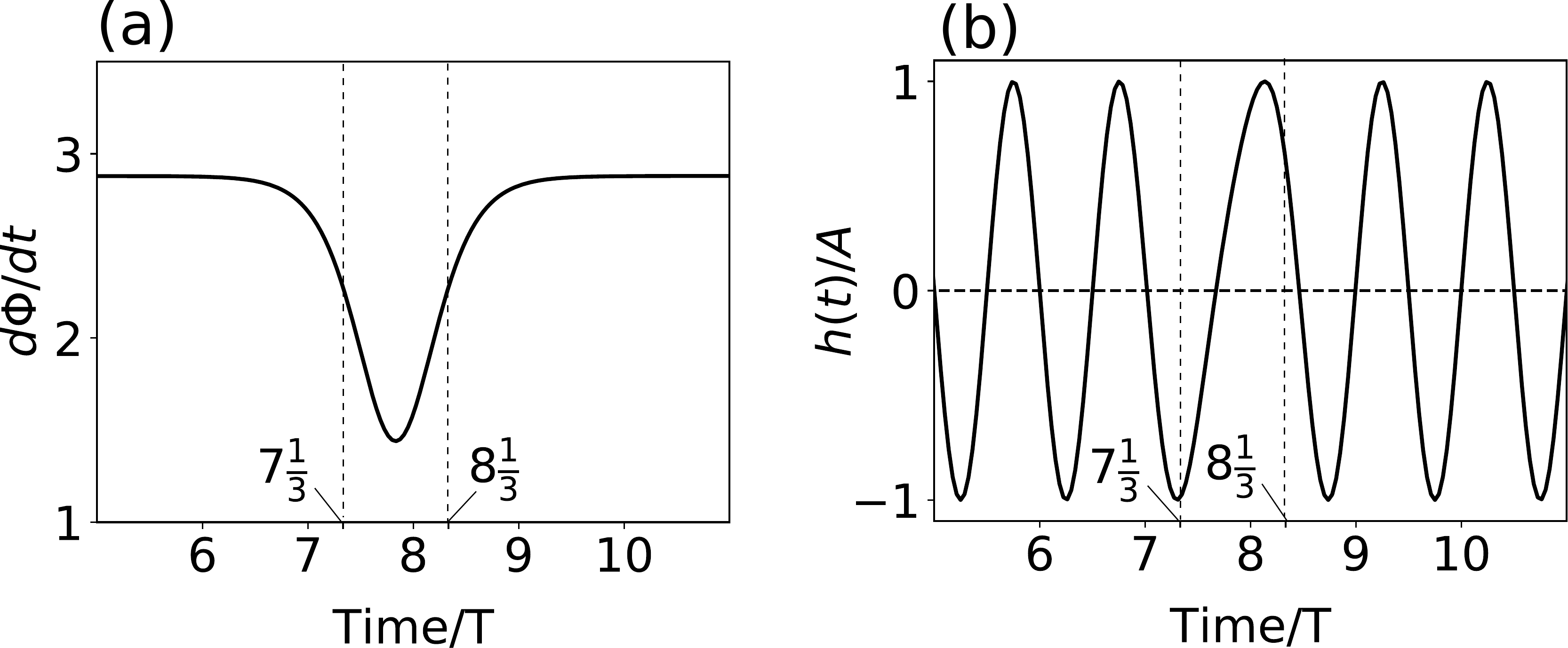}\
\caption{
Wing kinematics for $\Delta \omega=\omega/2, t_1=7\dfrac13 T, t_2=8\dfrac13 T (T=2.1818)$.
(a) $\displaystyle \dfrac{\partial \Phi(t)}{\partial t}$.
(b) $\frac{h(t)}{A}$.
}
\label{fig:wing motion}
\end{figure}

\begin{figure}
\includegraphics[width=0.9\textwidth]{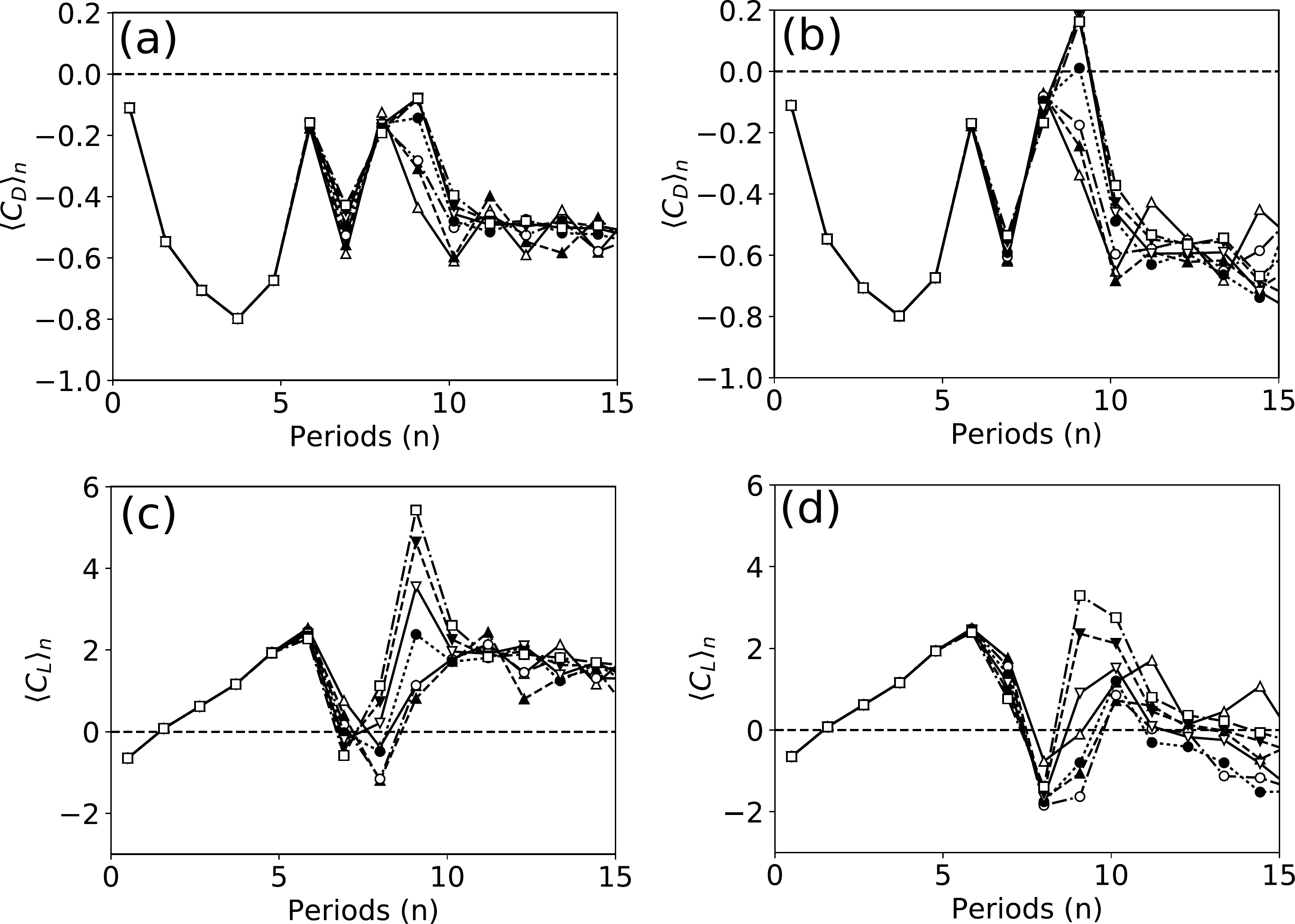}
 \caption{
(a) $\langle C_D \rangle_n$ vs. $n$ ($n$ is the period) for $t_1/T=7.0$.
 The open triangle($\vartriangle$),
 filled triangle($\blacktriangle$),
 open circle($\circ$),
 filled circle($\bullet$),
 open inverted triangle($\triangledown$),
 closed inverted triangle($\blacktriangledown$)
 and open square($\square$) indicate
 $(t_2-t_1)/T=0.4, 0.6, 0.8, 1.0, 1.2, 1.4, 1.6$, respectively.
(b) Same as (a), but $t_1/T=7\dfrac13$.
(c) $\langle C_L \rangle_n$ vs $n$ for $t_1/T=7.0$.
(d) Same as (c), but $t_1/T=7\dfrac13$.  
}
\label{fig:period averaged drag and lift}
\end{figure}

\begin{figure}
\includegraphics[width=0.8\textwidth]{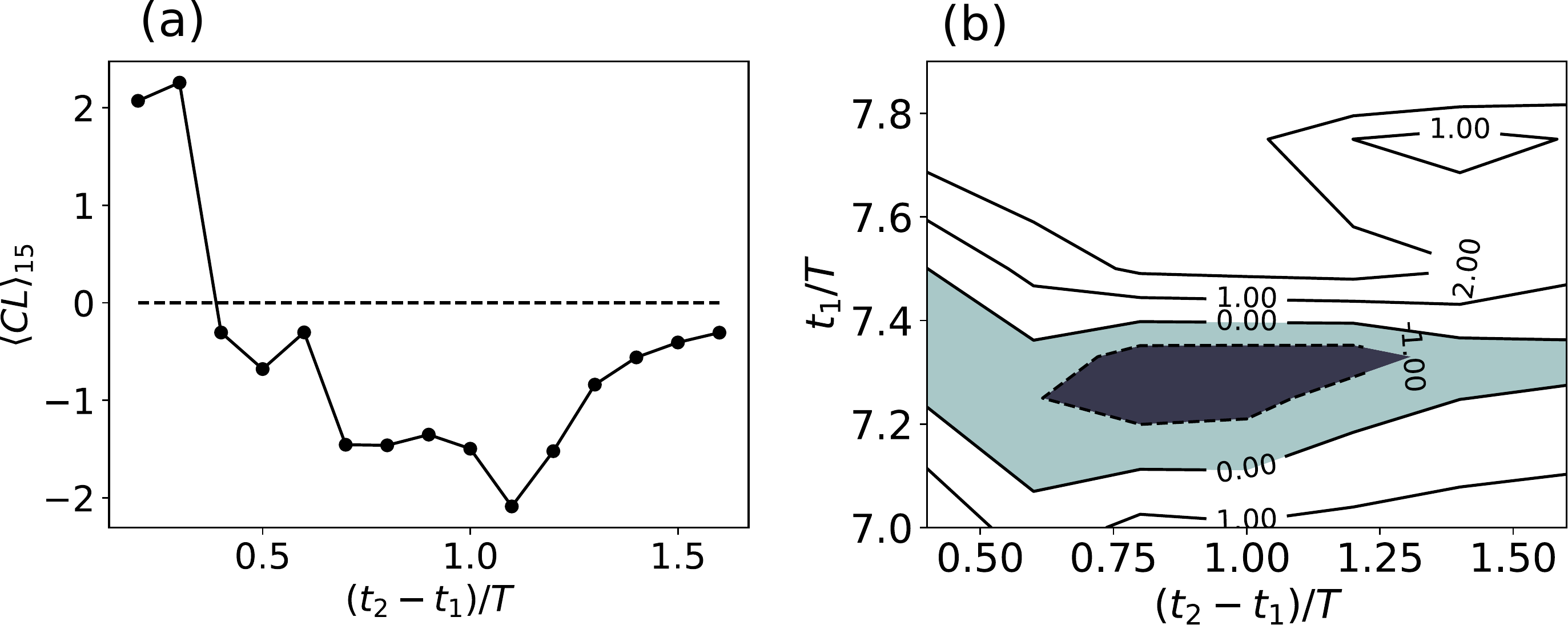}
\caption{
(a) $\langle C_L \rangle_{15}$ vs. $(t_2-t_1)/T$ for $t_1/T=7\dfrac13$.
(b) Contour of $\langle C_L \rangle_{15}$ for sets of $(t_1/T, (t_2-t_1)/T)$,
 where $t_1/T \in \{7, 7\dfrac14, 7\dfrac13, 7\dfrac12, 7\dfrac34, 7\dfrac56 \}$ and
 $(t_2-t_1)/T \in \{0.4, 0.6, 0.8, 1.0, 1.2, 1.4, 1.6\}$.
The region $\langle C_L \rangle_{15}<0$ is shaded.
}
\label{fig:averaged lift}
\end{figure}

\begin{figure}
\includegraphics[width=0.8\textwidth]{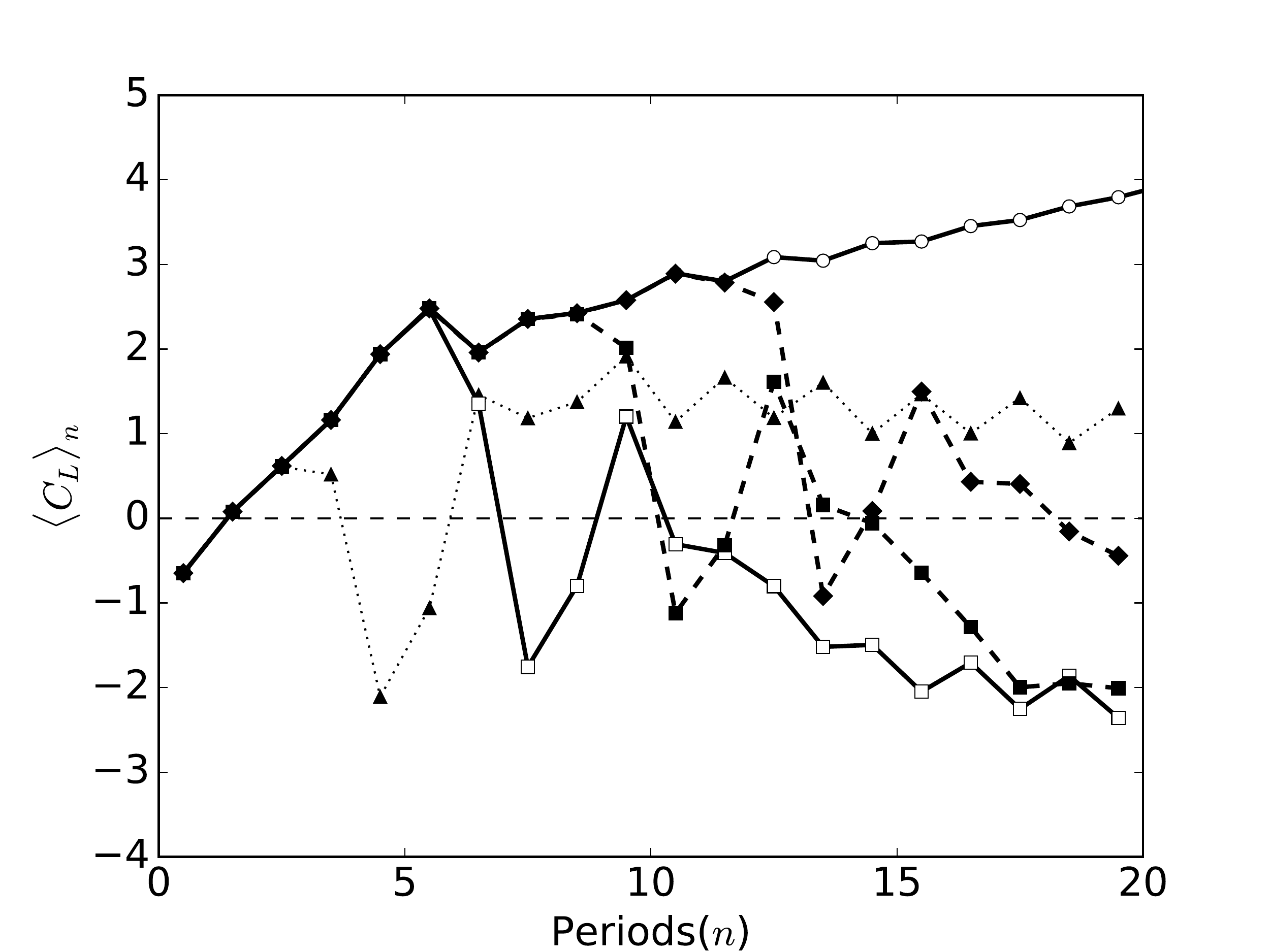}
\caption{
$\langle C_L \rangle_n$ vs. $n$.
The open circles indicate the case of no maneuvering ($\Delta \omega=0$).
The filled triangles($\blacktriangle$),
 open squares($\square$),
 filled squares($\blacksquare$)
 and
 diamonds ($\blacklozenge$) indicate
 $t_1/T=4\dfrac13, 7\dfrac13, 10\dfrac13$ and $13\dfrac13$, respectively.
}
\label{fig:long time period averaged lift}
\end{figure}

\begin{figure}
\includegraphics[width=0.9\textwidth]{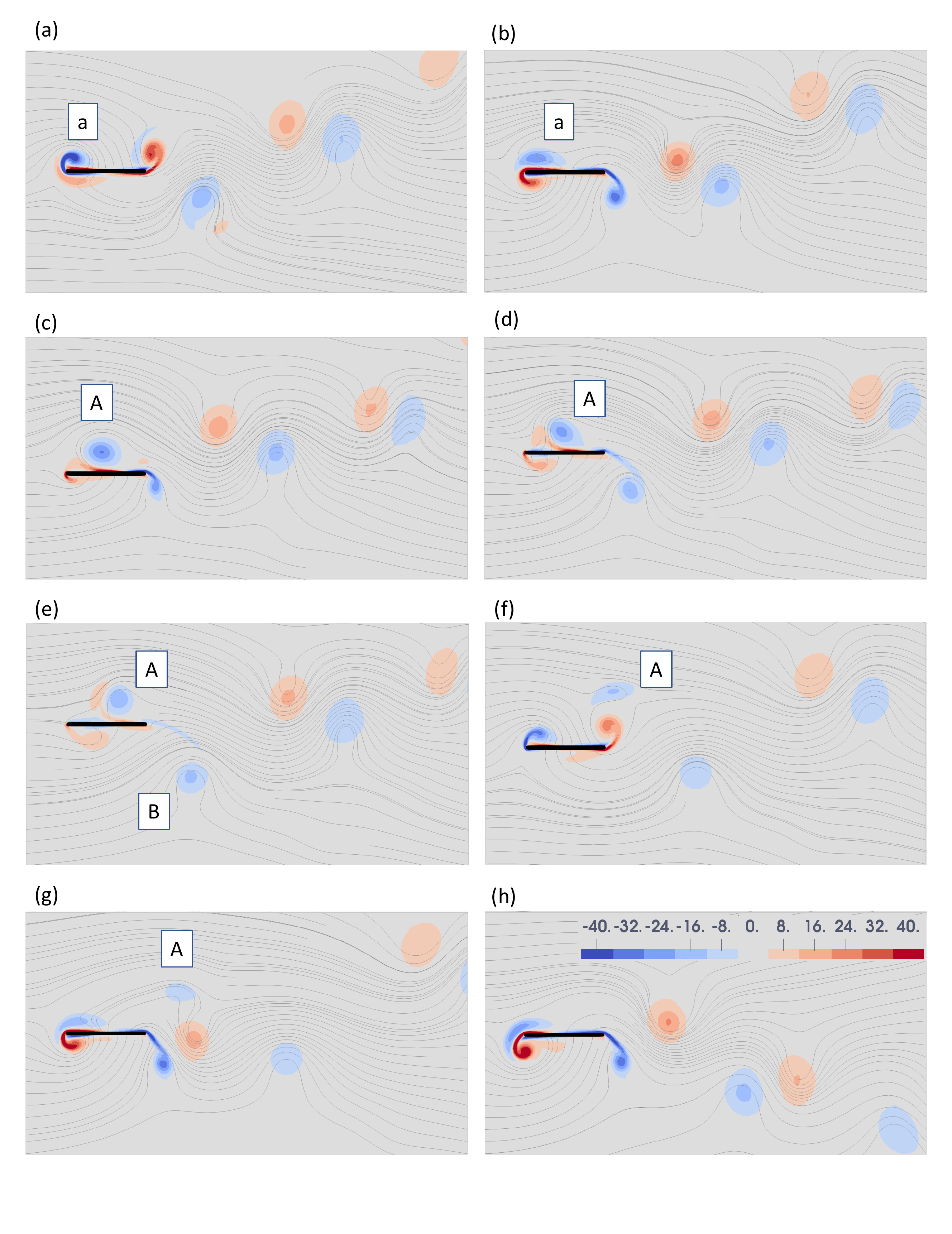}
\caption{
Vortex patterns during temporal reduction of the local angular frequency
 ($t_1/T=7\dfrac13, t_2-t_1=T$).
Curves indicate streamlines.
(a) $t=6.0T$.\;
(b) $t=6.5T$.\;
(c) $t=7.5T$.\;
(d) $t=7.75T$.\;
(e) $t=8.0T$.\;
(f) $t=8.5T$.\;
(g) $t=9.0T$.\;
(h) $t=14.0T$.
}
\label{fig:vortex patterns in flight maneouvre}
\end{figure}

\begin{figure}
\includegraphics[width=0.7\textwidth]{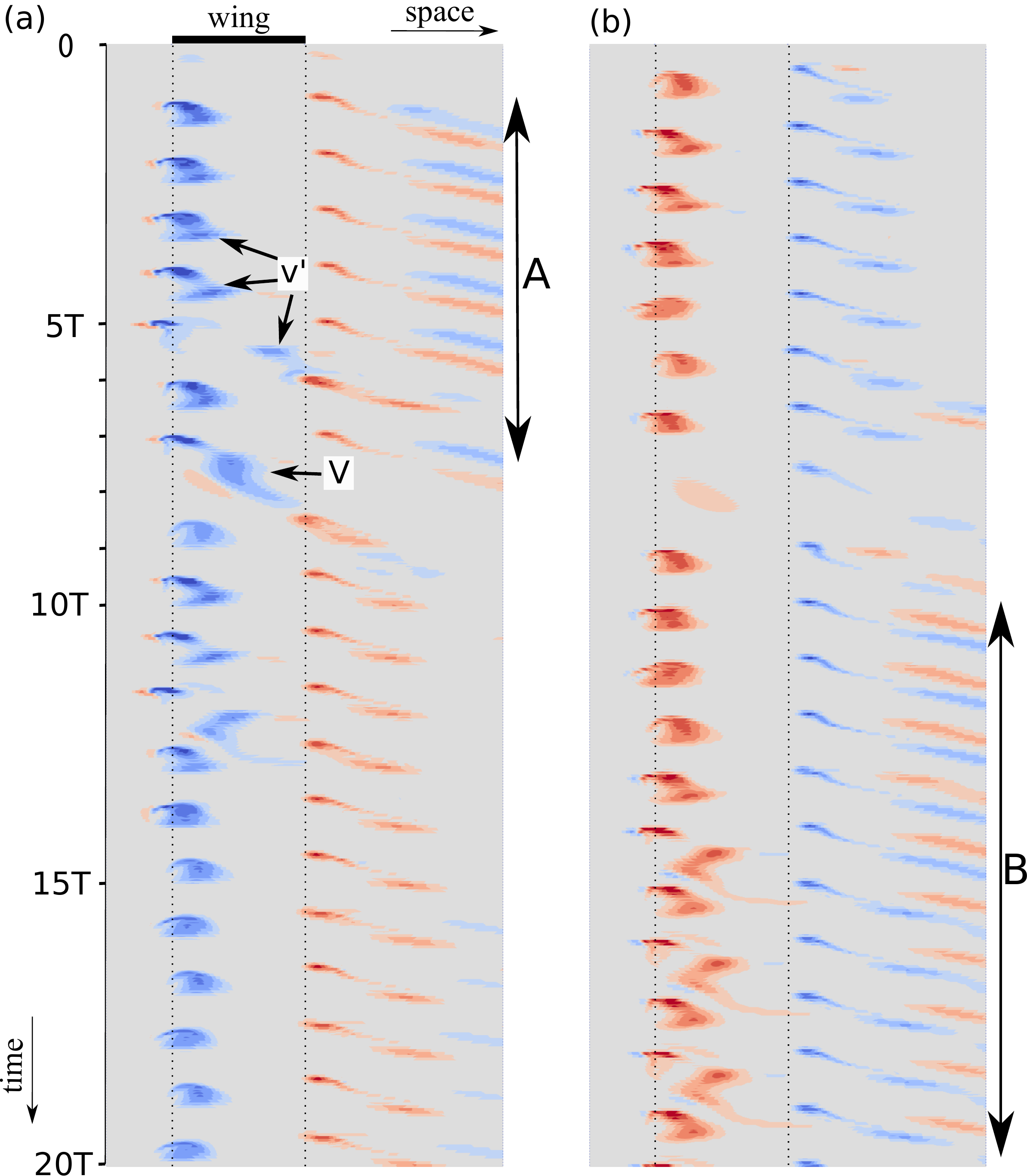}
\caption{
Time expanded image for $t_1/T=7\dfrac13, t_2-t_1=T$.
This image was generated by stacking the horizontal line distribution of vorticity from top to bottom,
 for the interval $0 \le t \le 20T$, $-2 \le x \le 4$.
The color legend is the same as that for Fig. \ref{fig:Leading-edge vortex for St0.275}.
(a) image for the horizontal line $c/4$ above the wing.
(b) image for the horizontal line $c/4$ below the wing.
}
\label{fig:time expaneded picture}
\end{figure}

In this section, we demonstrate that an inversion of the vortex pattern associated with the lift sign can be triggered by a temporal reduction in the heaving frequency when $St=0.275$, $\phi=\pi$, and $\Delta \omega = \omega/2$. 
Figure \ref{fig:wing motion} shows the local angular frequency $\partial \Phi/\partial t$ and non-dimensional heaving motion $h(t)/A$ for $(t_1/T, (t_2-t_1)/T)=(7\dfrac13, 1)$. 
In this case, the frequency reduction is apparent in the eighth flapping, while the change is smooth according to the definition of the class of the wing motion.

Figure \ref{fig:period averaged drag and lift} shows series of $\langle C_L \rangle_n$ and $\langle C_L \rangle_n$ for two typical cases, $t_1/T=7$ and $7\dfrac13$,
 to demonstrate how the inversion depends on $(t_1, t_2)$. 
In both the cases, a strong disturbance due to the wing motion causes a significant change in the period-averaged values. The results for the case $t_1/T=7$ are shown in Figs. \ref{fig:period averaged drag and lift}(a) and (c). 
In this case, regardless of the value of $t_2$, the values of $\langle C_L \rangle_n$ and $\langle C_D \rangle_n$ in the last period did not change significantly. 
In particular, the sign of $\langle C_L \rangle_n$ did not change for all values of $t_2$ in $0.4 \le (t_2-t_1)/T \le 1.6$. 
In other words, the disturbances when $t_1/T=7$ did not change the eventual vortex structures.

The results for the case $t_1/T=7\dfrac13$, in which $t_1$ is slightly different value from that of the above-mentioned case, are shown in Figs. \ref{fig:period averaged drag and lift}(b) and (d). It is clear that the sign of the lift is inverted for several values of $t_2$ with decreasing values in the latter periods. 
Figure \ref{fig:averaged lift}(a) shows the detailed values of $\langle C_L \rangle_{15}$ for $t_1/T=7\dfrac13$. 
A wide range of values of $(t_2-t_1)/T$ gives greater negative values (e.g., $\langle C_L \rangle_{15}<-1$ for $0.7 \le (t_2-t_1)/T \le 1.2)$). 
These results suggest that the lift inversion is robust for the values of $(t_2-t_1)$ around $(t_2-t_1)/T=1.0$.

The result of a parametric study on the lift inversion is shown in Fig. \ref{fig:averaged lift}, where $\langle C_L \rangle_{15}$ was used. In Fig. \ref{fig:averaged lift}(b), the contour of $\langle C_L \rangle_{15}$, which was used as a characteristic variable of the inversion, is shown for sets of $(t_1/T, (t_2-t_1)/T)$. We can see that the range around $(t_1/T, (t_2-t_1)/T)=(7\dfrac13,1)$ gives negative values (shaded region in Fig. \ref{fig:averaged lift}(a)) corresponding to the lift inversion.
We remark that the time interval of reduced frequency $t_2-t_1$ considered here is less than $1.5T$, which implies that the phenomena discussed here is due to unsteady (non-periodic) dynamics of wing and the flow.

Next, the robustness to the characteristic start time of the frequency reduction, $t_1$, is examined. Figure \ref{fig:long time period averaged lift} shows $\langle C_L \rangle_n$ as a function of $n$. 
We fixed $(t_2-t_1)/T=1$. 
Four cases for $t_1/T=4\dfrac13, 7\dfrac13, 10\dfrac13$ and $13\dfrac13$ and a case of no frequency reduction $(t_1 \to \infty)$ are shown to examine the relationship between the vortex structure at the beginning of the reduction and the final result. 
Note that the phase of the frequency reduction was fixed because it is important for the inversion, as shown in Figure \ref{fig:averaged lift}(a). Figure \ref{fig:long time period averaged lift} shows that the case $t_1/T=4\dfrac13$ did not reach the final inversion, which suggests that the vortex structure generated up to this time is not sufficiently ``mature'' to accept the transition mechanism discussed below. 
The time series of $\langle C_L \rangle_n$ for $t_1/T=7\dfrac13, 10\dfrac13, 13\dfrac13$ are rather universal. 
The lift sign changes once to negative during the temporal frequency reduction and then changes to positive for one or two periods; finally, decreasing negative values are observed. 
This result suggests that the process of the transition in the vortex structure has a universality property and the temporal frequency reduction strategy for the lift inversion requires a ``matured'' vortex structure that contains several coherent vortices. 
In the following, we consider the case
 $t_1/T=7\dfrac13, (t_2-t_1)/T=1$
 as a typical example.
We note that similar inversion process is observed when $0.26 \le St \le 0.28$ while keeping other parameters the same.

The vortex dynamics during the reduction is shown in Fig. \ref{fig:vortex patterns in flight maneouvre} (See Supplemental Material for [Re200Maneuver.mp4]). 
As explained in Sec. \ref{sec: Simple heaving}, the LEVs are dissipated or trapped near the leading edge in the simple heaving motion, which also occurs before the reduction starts (``a'' in Figs. \ref{fig:vortex patterns in flight maneouvre}(a) and (b)). 
However, the temporal frequency reduction weakens the stretching or dissipation process of the LEVs. Because of slower upward motion of the wing in this process, the LEV above the wing did not stretch considerably, resulting in its survival (``A'' in
 Fig. \ref{fig:vortex patterns in flight maneouvre}(c)). 
Because this vortex is generated before the frequency reduction, the magnitude of the vorticity is close to the corresponding vortex in simple heaving
(Figs. \ref{fig:CL for Long time}(b) and (c)), although the detailed shape depends on the wake structure.
Figs. \ref{fig:vortex patterns in flight maneouvre}(c)-(f) shows the dynamics during the frequency reduction interval.
The streamlines indicate that the flow around the LEV ``A'', which is detached from the leading edge, is rightward.
The time interval of the frequency reduction is sufficient
 to transfer the LEV ``A'' to the trailing edge, and
 the LEV reaches without significant distortion or dissipation.
The negative sign of the LEV induces stronger local velocity near the trailing edge. As a result, the LEV shifts the position of the TEV generated in this interval (``B'' in
Fig. \ref{fig:vortex patterns in flight maneouvre}(e)). Moreover, the LEV ``A'' remains near the trailing edge during the next period to induce a backward flow so that the vortices near the trailing edge are not advected excessively. Then, a dipole vortex moving in an obliquely downward direction is generated (Figs. \ref{fig:vortex patterns in flight maneouvre}(f) and (g)). The arrangements of the vortices significantly change the position of the subsequently generated coherent vortices to finally invert the vortex pattern (Fig. \ref{fig:vortex patterns in flight maneouvre}(h))
. 
The vortex patterns in 
Figs. \ref{fig:vortex patterns in flight maneouvre}(a) and (h) are roughly symmetric with respect to a horizontal line (note that the phase of the wing oscillation shifted by $-\Delta \omega (t_2-t_1)=-\pi$).

It should be remarked that the unsteady wing-vortex interaction during the lift inversion matches the successful parameters of  $(t_1, t_2)$. Actually, Fig.\ref{fig:averaged lift}(b) suggests that $(t_2-t_1)/T$ should be around unity and the phase of $t_1$ is around $1/3$ (of the period). The former corresponds to the order of time to transfer LEV along the wing cord, and the latter is a condition for the wing to upward slowly during frequency reduction period.

The essential part of this transition dynamics can be extracted from the time-expanded images shown in Fig. \ref{fig:time expaneded picture}. 
Figure \ref{fig:time expaneded picture}(a) is generated by stacking the line vorticity distribution $c/4$ above the wing;
 a horizontal cross section of the figure shows the spatial vorticity distribution and a vertical cross section shows the time series of the vorticity at a particular point. 
In the time interval ``A'', both red and blue lines are shown in turn on the right (downstream). 
These lines indicate positive and negative vortices generated to form the deflected wake with a positive deflection angle (corresponding to Fig. \ref{fig:vortex patterns in flight maneouvre}(a)). 
The LEV transfer above the wing is indicated by the blue region ``V'' for the time interval $[7T, 8T]$. The transferred LEV interacts with the TEV to change the deflection angle. After the interaction, the blue lines, corresponding to negative vortices, disappear because the deflection angle is inverted.

Figure \ref{fig:time expaneded picture}(b) is generated similarly to Fig. \ref{fig:time expaneded picture}(a), but for the line vorticity distribution $c/4$ \textit{below} the wing. 
The vortex pattern before the frequency reduction is similar to that in Fig. \ref{fig:time expaneded picture}(a) after the frequency reduction with the inverted sign, suggesting that the deflection pattern is inverted during the process. Similarly, the vortex pattern in Fig. \ref{fig:time expaneded picture}(a) during the period ``A'' is similar to that in Fig. \ref{fig:time expaneded picture}(b) during the period \textit{after} the frequency reduction with the inverted sign (indicated by ``B''). These images clearly show the inversion dynamics, especially for the effect of LEV transfer. 

Furthermore, the vortex pattern indicated by ``v'' shows an irregular transfer of a part of the LEV without temporal frequency reduction. 
Although the entire vortex pattern is disturbed by such an irregular vortex transfer that is observed sometimes, it does not change significantly (cf. Fig. \ref{fig:CL for Long time}).

\subsection{Discussion: The Reynolds number dependency}
\label{sec: Discussion: The Reynolds number dependency}
\begin{figure}[h]
\includegraphics[width=0.8\textwidth]{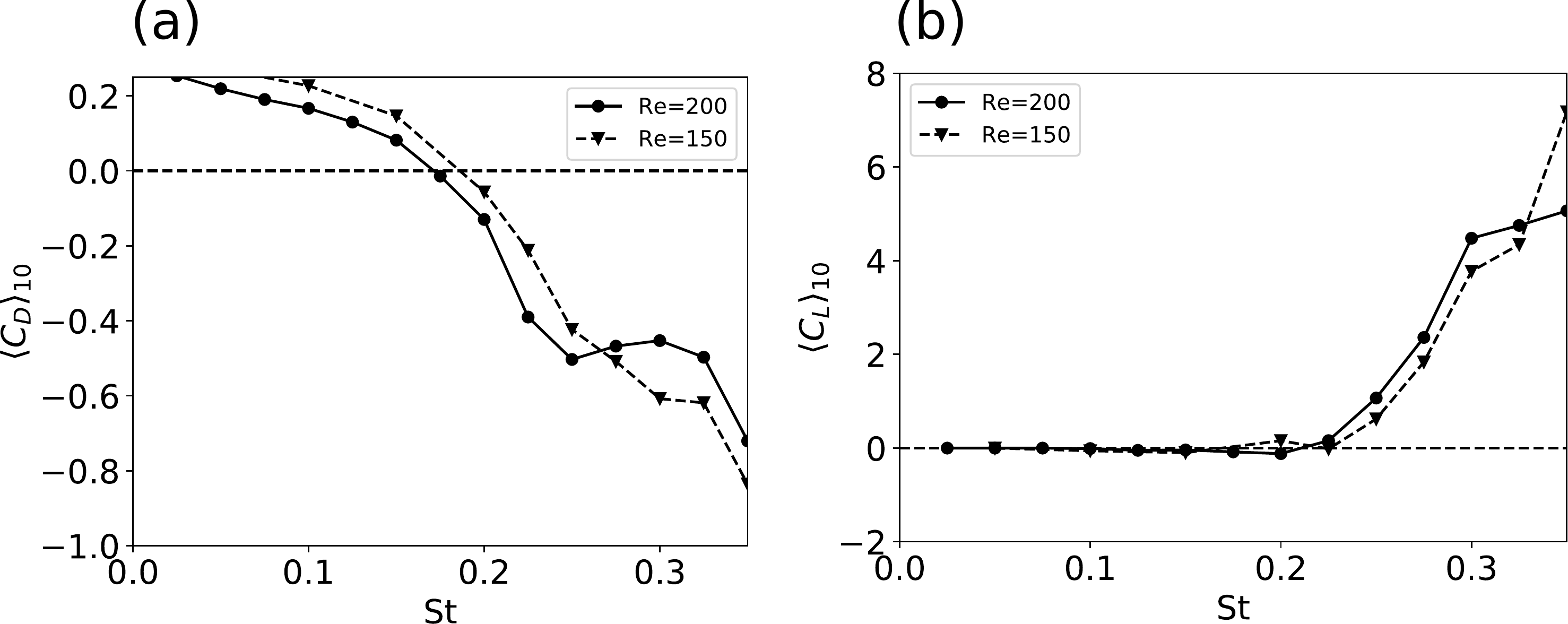}
\caption{
(a) Reynolds number dependence on the transitions of $\langle C_D \rangle_{10}$.
(b) Same as (a) but for $\langle C_L \rangle_{10}$.
}
\label{fig:CDL_Re}
\end{figure}

\begin{figure}[h]
\includegraphics[width=0.7\textwidth]{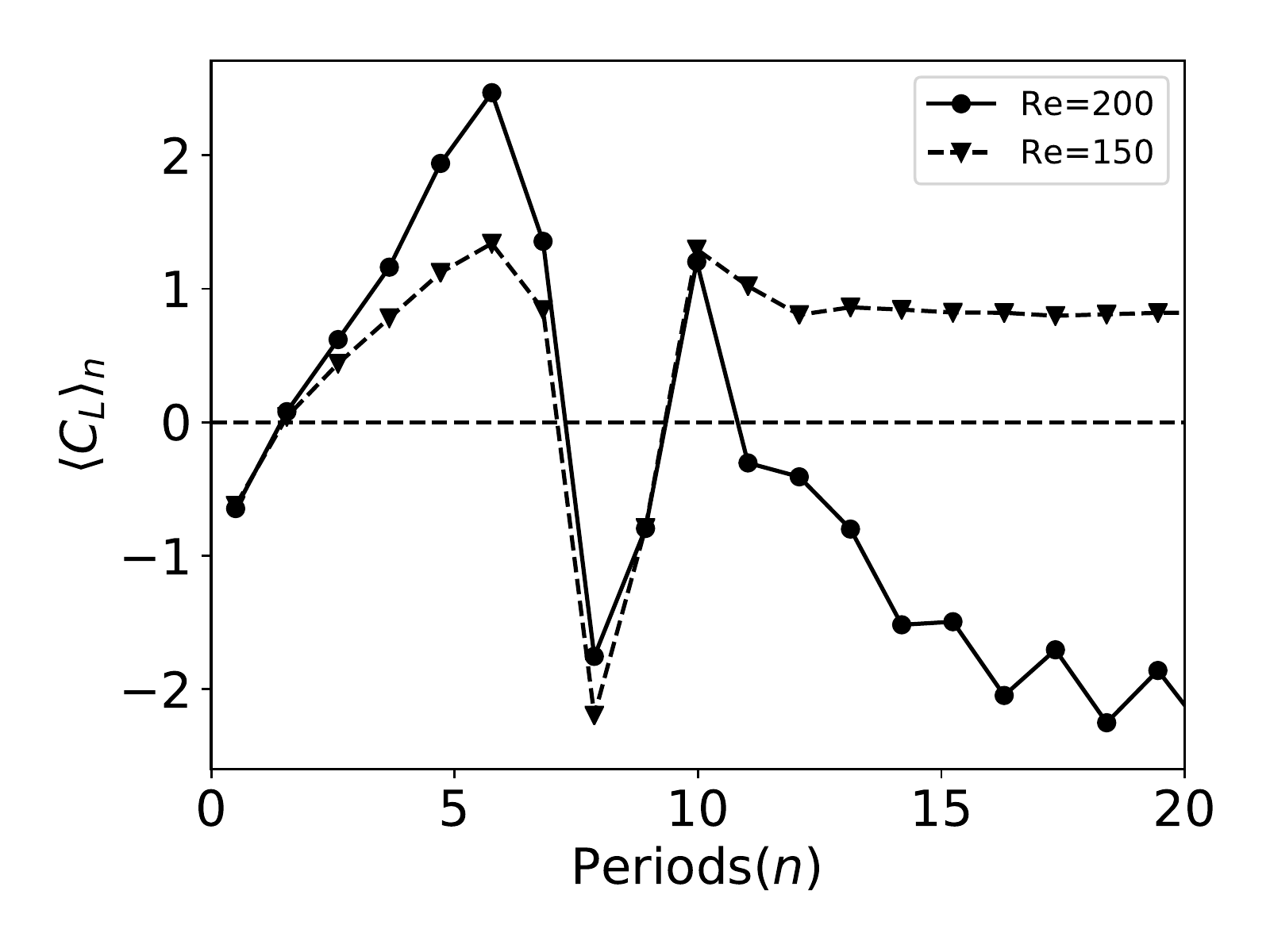}
\caption{
$\langle C_L \rangle_n$ vs $n$ for the cases $Re=150, 200$, where $t_1/T=7\dfrac13$ and $t_2-t_1=T$.
}
\label{fig:CL_Maneuver_Re}
\end{figure}

In this subsection, we discuss the Reynolds number dependence on the transitions of the vortex structures.

First, the transition behaviors of $\langle C_D \rangle_n$ and $\langle C_L \rangle_n$ for different $St$ values are determined for the case $Re=150, \phi=\pi$. 
Figure \ref{fig:CDL_Re} shows the result, together with the case $Re=200$, which suggests that the vortex transition for different $St$ values does not change significantly in this interval, though a slight increment in the value of $\langle C_D \rangle$ is observed for $Re=150$. 
However, in particular, the behavior at $St=0.275$, which has been discussed in detail in Sec. \ref{sec: Transition of vortex pattern}, is the same, i.e., thrust is generated and the asymmetric vortex pattern is observed. 
The transition behaviors of the lift and the vortex structures for the temporal frequency reduction were compared for the typical case: $t_1/T=7\dfrac13, (t_2-t_1)/T=1$. 
Figure \ref{fig:CL_Maneuver_Re} shows the result; clearly, the lift inversion fails when $Re=150$.

The difference is attributed to the large dissipation. 
A long-time simulation of simple heaving shows that $\langle C_L \rangle_n$ attains a plateau $t>90T$ for the case $Re=150$ (data not shown); by contrast, for the case $Re=200$, $t\simeq 50T$ (Fig. \ref{fig:CL for Long time}). 
Moreover, the initial increase rate of $\langle C_L \rangle_n$ for $Re=150$ is smaller than that for $Re=200$. 
Because the number of vortices in the vortex structures at a particular time does not depend on $Re$, the low increase rate is attributed to the large dissipation. 
In this case, the LEV is transferred as in the case of $Re=200$, but there is no rearrangement of the TEV that leads to the inversion of the lift or the vortex structure. We performed a similar analysis for the cases $Re=170$ and $Re=180$, and we found that the critical Reynolds number seems to lie between $Re=170$ and $Re=180$ (data not shown).

\section{Concluding Remarks}
\label{sec:Concluding Remarks}

In this paper, we studied the inversion of the lift and the asymmetric vortex pattern of a heaving wing in a uniform flow under a temporal reduction of the angular frequency.
In a parameter range, it is possible to invert the vortex pattern and the lift sign. During the inversion process, the LEV plays an important role. Without the temporal reduction of the local angular frequency, the LEV dissipates or remains near the leading edge, and it does not contribute to the vortex generation at the trailing edge significantly. However, during the temporal reduction of the local angular frequency, the LEV is advected to the trailing edge to enhance the local flow, which triggers the inversion process initiated by the position shift of the generated TEV.

We demonstrated that it is possible to control the vortex structure via the wing motion, but such control is not straightforward even in our simple configuration. Previously considered examples of lift generation based on vortex generation include wake capture \cite{dickinson99_wing_rotat_aerod_basis_insec_fligh} and the symmetry-breaking mechanism of symmetric flapping models \cite{%
iima01_is_two_dimen_butter_able,%
ota12_lift_gener_by_two_dimen%
}. However, the mechanism presented here is used to change the qualitative vortex structure, which is different from the above-mentioned mechanisms.

Efficient usage of the LEV-TEV interference can lead to vortex pattern inversion. As discussed in Sec. \ref{sec: Simple heaving}, suppression of the LEV-TEV interference is required to maintain the deflected wake under regular flapping. 
Such interference might be exploited for lift vector control in the future.

It is interesting to note that the Strouhal number in the flight and swimming of many animals lies in the range of $0.2-0.4$\cite{taylor03_flyin_swimm_animal_cruis_at}, and the authors suggest that the vortex pattern generated in this range is a key underlying factor. In this region, the LEV is shed as the downstroke ends, which is in agreement with our result that the LEV transfer causes the vortex pattern change. Our results suggest that such LEV shedding behavior might be useful not only for maintaining flapping flight but also for maneuverability. Although our results are restricted to the transition of the vortex dynamics owing to change of the wing motion, we believe that they will facilitate a deeper understanding of the maneuverability of flying animals.

\begin{acknowledgments}
This work was partially supported by JSPS KAKENHI Grant Number JP16H04303.
\end{acknowledgments}

\appendix*
\section{Validation}
\label{sec:Appendix: Validation}
\begin{figure}[h]
\includegraphics[width=0.8\textwidth]{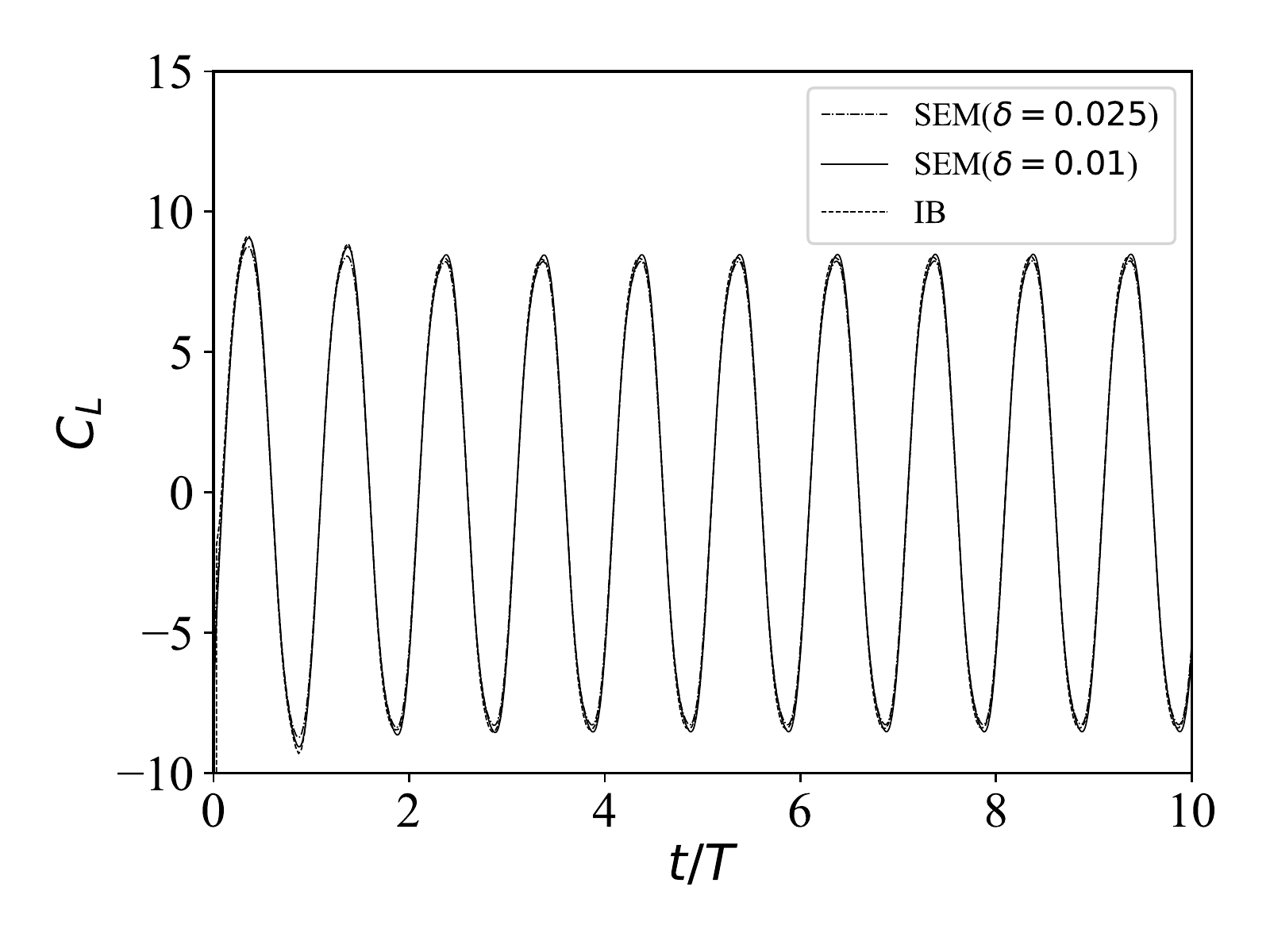}
\caption{
Time series of the lift on the oscillating wing in a uniform flow. 
The calculation methods are SEM ($\delta=0.025, 0.01$) and IB.
}
\label{fig:comparison with Yokoyama}
\end{figure}

The validity of the simulation code and the algorithm for the transformation between the laboratory frame ($O-xy$) and the wing-fixed frame ($O-XY$) was verified by comparing the lift acting on the oscillating wing in a uniform flow. We compared the SEM code with the code of the immersed boundary (IB) method, which is a variant used by Yokoyama et al. \cite{yokoyama13_aerod_forces_vortic_struc_flapp}. In the calculation of the IB method, the computational domain was $[-10,30]\times[-5,5]$, which was represented by $1024 \times 256$ modes corresponding to the regular intervals. The time step for the IB method was $1.6 \times 10^{-5}$. The wing chord was represented by 64 grid points, corresponding to $c=2.5$. We compared our SEM approach with the IB approach for the case $Re=200, St=0.15, r=0.2$. 
Figure \ref{fig:comparison with Yokoyama} shows the lift coefficient $C_L$ calculated by both methods. The time series are nearly identical, especially for the case $\delta=0.01$.

\end{document}